\documentclass{INTERSPEECH2023}

\usepackage{booktabs, multirow}
\usepackage{amssymb}
\usepackage{caption}
\usepackage{subcaption}

\interspeechcameraready

\title{Diverse and Expressive Speech Prosody Prediction with Denoising Diffusion Probabilistic Model}
\name{Xiang Li$^{1,\dagger}$,\thanks{$\dagger$ Work completed during internship in Tencent AI Lab} Songxiang Liu$^{2}$, Max W. Y. Lam$^{2}$, Zhiyong Wu$^{1,3,*}$\thanks{* Corresponding author.}, Chao Weng$^{2}$, Helen Meng$^{1,3}$}
\address{
  $^1$Tsinghua-CUHK Joint Research Center for Media Sciences, Technologies and Systems,\\Shenzhen International Graduate School, Tsinghua University, Shenzhen, China\\
  $^2$Tencent AI Lab, Shenzhen, China\\
   $^3$Department of Systems Engineering and Engineering Management,\\The Chinese University of Hong Kong, Hong Kong SAR, China}
\email{xiang-li20@mails.tsinghua.edu.cn, \{shaunxliu, maxwylam\}@tencent.com, zywu@sz.tsinghua.edu.cn, cweng@tencent.com, hmmeng@se.cuhk.edu.hk}

\begin{document}

\maketitle

\begin{abstract}
Expressive human speech generally abounds with rich and flexible speech prosody variations.
The speech prosody predictors in existing expressive speech synthesis methods mostly produce deterministic predictions,
which are learned by directly minimizing the norm of prosody prediction error.
Its unimodal nature leads to a mismatch with ground truth distribution and harms the model's ability in making diverse predictions.
Thus, we propose a novel prosody predictor based on the denoising diffusion probabilistic model to take advantage of its high-quality generative modeling and training stability.
Experiment results confirm that the proposed prosody predictor outperforms the deterministic baseline on both the expressiveness and diversity of prediction results with even fewer network parameters.
\end{abstract}
\section{Introduction}
\label{sec:intro}
Speech synthesis aims to convert a given text into the corresponding speech audio.
In modern speech synthesis models,
the naturalness and sound quality of the synthesized audios have been promoted to a human-like level by incorporating neural-network-based acoustic models and vocoders,
which are now widely accessible in many real-world applications \cite{Wang2017,shen2018natural,ren2021fastspeech,DBLP:conf/nips/KongKB20}.
On the other hand, how to reach the expressiveness and diversity of human speech in synthesized audio remains an open question.
Thus, the task of expressive speech synthesis is drawing growing attention recently.

Existing expressive speech synthesis methods generally model the prosody in human speech as an additional condition to the text-to-speech (TTS) backbone to stylize the synthesized audio
\cite{skerry2018towards,lee2019robust,li21r_interspeech}.
During training, the speech prosody representation is extracted from the target speech as the desired condition of the TTS backbone and the training target of a speech prosody predictor,
which makes its prediction according to the input text \cite{hodari2021camp}.
During inference, given the input text, the prediction result of the trained speech prosody predictor is applied as the actual condition of the TTS backbone.
As a result, the speech prosody prediction module is not only necessary for the inference process in real-world applications, but also the crucial component that directly affects the expressiveness and diversity of the prosody in synthesized speech.

Typical speech prosody predictors like the phoneme duration, fundamental frequency (pitch), and energy predictors in FastSpeech2 \cite{ren2021fastspeech} and FastPitch \cite{lancucki2021fastpitch} take pre-extracted duration, pitch, and energy features as the target speech prosody representation, and utilize the output of the text encoder in TTS backbone as prediction conditions.
Other methods manage to improve the prediction conditions by incorporating context information or semantic structure of the input text \cite{ren2022prosospeech,lei2022towards}.
The prosody predictors in these methods 
generally establish a deterministic mapping from input text to predicted prosody,
and is trained via minimizing the L1 or L2 norm of prediction error,
by assuming that the target prosody distribution is an independent unimodal Laplace or Gaussian distribution \cite{ren2022revisiting}.

However, the prosody of human speech is extremely flexible.
Changes in intonation, speaking rate, emphasis, context, and other factors provide numerous ways to utter the same text,
resulting in multimodal and correlated prosody distributions.
Consequently, the diversity and expressiveness of the prediction results from previous prosody predictors tend to be diminished by its deterministic characteristics and the over-smoothing effect due to simplified assumptions on prosody distribution.

Thus, we propose to replace the deterministic prosody predictors with an alternative based on the denoising diffusion probabilistic model (DDPM) \cite{ho2020denoising}.
DDPM and its variants have shown powerful capacity in generative modeling with respect to high-quality sampling, gorgeous model coverage, and sample diversity \cite{dhariwal2021diffusion,DBLP:conf/iclr/KongPHZC21,jeong21_interspeech,DBLP:journals/corr/abs-2201-11972,10.1007/978-3-031-15063-0_36}.
Moreover, DDPMs can be trained with a simplified variant of the evidence lower bound of data likelihood, leading to stable optimization.
Based on this thinking,
this paper focuses on investigating the use of DDPMs for modeling the diverse and flexible one-to-many mapping from text to prosody in human speech.
Experiment results reveal that the proposed DDPM-based prosody predictor significantly outperforms the deterministic baseline on both the diversity and expressiveness of prediction results,
indicating the effectiveness of introducing DDPM to the task of speech prosody prediction.


\begin{figure*}[th]
\centering
\begin{minipage}[b]{.45\textwidth}
  \centering
  \includegraphics[width=0.5\textwidth]{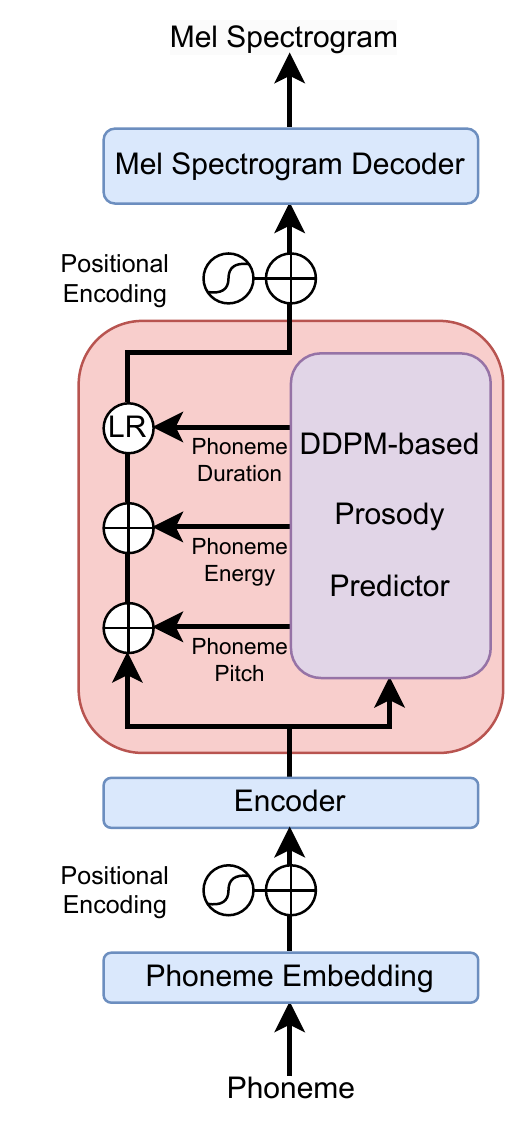}
  \subcaption{The FastSpeech2-based TTS backbone}
  \label{fig:model_all}
\end{minipage}%
\begin{minipage}[b]{.55\textwidth}
  \centering
  \includegraphics[width=.95\textwidth]{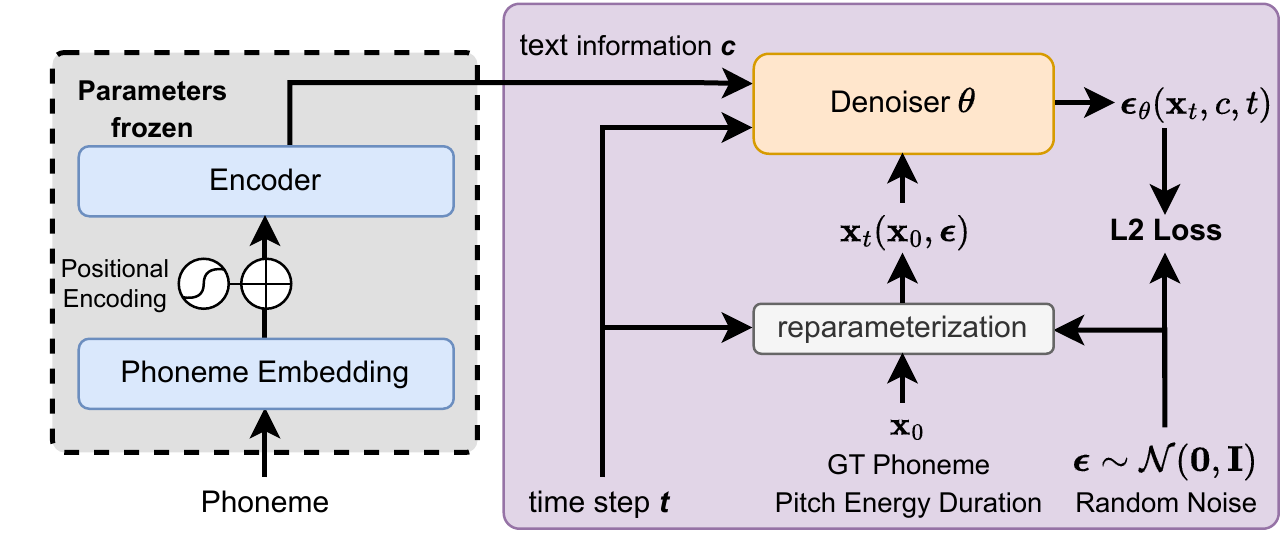}
  \subcaption{The training procedure of the DDPM-based prosody predictor}
  \label{fig:diffvar_train}
  \includegraphics[width=0.9\textwidth]{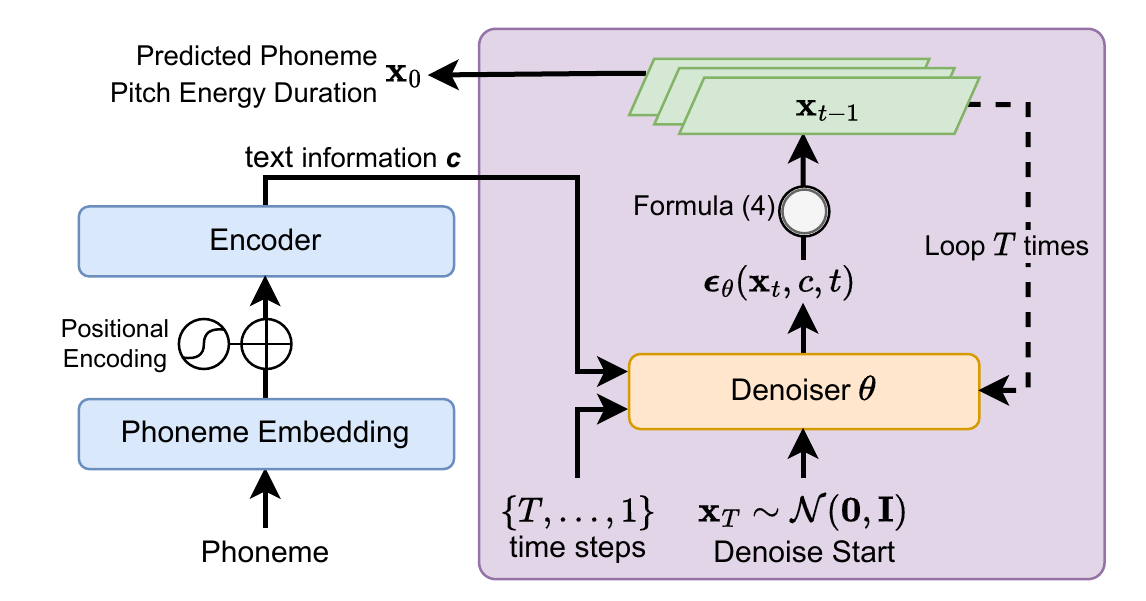}
  \subcaption{The inference procedure of the DDPM-based prosody predictor}
  \label{fig:diffvar_infer}
\end{minipage}
\caption{The overall architecture of the expressive TTS system with the proposed DDPM-based prosody predictor}
\label{fig:model}
\vspace{-0.5cm}
\end{figure*}
\section{Methodology}
\label{sec:method}

As shown in Figure~\ref{fig:model}, we incorporate the proposed DDPM-based prosody predictor with a pre-trained TTS backbone based on FastSpeech2 to construct an expressive TTS system.
Following FastSpeech2, we encode speech prosody into a 3-dimensional feature of pitch, energy, and duration.

\subsection{Text-to-speech backbone}
\label{sec:method-backbone}
The backbone of the proposed expressive TTS system is an adapted FastSpeech2 with a modified variance adaptor.
The phoneme embedding, encoder, and mel spectrogram decoder modules are the same as those in FastSpeech2.
While for the variance adaptor,
instead of applying frame-wise pitch and energy features,
we use phoneme-wise pitch and energy to improve training stability.
The phoneme-wise pitch, energy, and duration values are provided by the proposed DDPM-based prosody predictor,
as displayed in Figure~\ref{fig:model_all}.
Similar to the original setting, the given pitch and energy are first quantized and mapped to embedding vectors via a codebook,
then added to the phoneme-wise output of the encoder as a stylized input condition,
which is repeated with a length regulator (LR) module according to the given phoneme durations and fed to the mel spectrogram decoder to generate the final mel spectrogram.

\subsection{DDPM-based prosody predictor}
\label{sec:method-diffvar}

\subsubsection{Preliminary knowledge on DDPM}
DDPM is a generative model built upon a forward process that is fixed to a Markov chain that diffuses data $\mathbf{x}_0$ into white noise, and a reverse process that samples $\mathbf{x}_0$ from white noise via another Markov chain with learned Gaussian transitions.

Practically, the former process 
diffuses the $\mathbf{x}_{t-1}$ with a small Gaussian noise to obtain $\mathbf{x}_{t}$ at each diffusion step $t\in[1,T]$ according to the fixed variance schedule $\beta_1,\dots,\beta_T$:
$$q(\mathbf{x}_t|\mathbf{x}_{t-1}):=\mathcal{N}(\mathbf{x}_t;\sqrt{1-\beta_t}\mathbf{x}_{t-1}, \beta_t\mathbf{I})$$
After $T$ steps, $\mathbf{x}_0$ is transformed into $\mathbf{x}_T$.

The reverse process starts at $p(\mathbf{x}_T)=\mathcal{N}(\mathbf{x}_T;\mathbf{0},\mathbf{I})$,
and is approximated via a neural denoiser $\theta$ as:
$$p_\theta(\mathbf{x}_{t-1}|\mathbf{x}_t):=\mathcal{N}(\mathbf{x}_{t-1};\boldsymbol{\mu}_\theta(\mathbf{x}_t,t),\sigma^2_t\mathbf{I})$$
where $\sigma^2=\frac{1-\overline{\alpha}_{t-1}}{1-\overline{\alpha}_t}\beta_t$, with $\alpha_t:=1-\beta_t$ and $\overline{\alpha}:=\prod^t_{s=1}\alpha_s$.
The parameters $\theta$ of the denoiser are trained via minimizing a variational bound of the negative log-likelihood:
$$\mathbb{E}[-\log p_\theta(\mathbf{x}_0)]\leq\mathbb{E}_q\left[-\log\frac{p_\theta(\mathbf{x}_{0:T})}{q(\mathbf{x}_{1:T}|\mathbf{x}_0)}\right]$$
which is eventually simplified as the following variant by the reparameterizations in Formula \ref{equation:reparam1} and \ref{eqaution:reparam2} where $\boldsymbol{\epsilon}\sim\mathcal{N}(\mathbf{0},\mathbf{I})$.
$$L_{simple}(\theta):=\mathbb{E}_{t,\mathbf{x}_0,\boldsymbol{\epsilon}}\left[\Vert\boldsymbol{\epsilon}-\boldsymbol{\epsilon}_\theta(\sqrt{\overline{\alpha}_t}\mathbf{x}_0+\sqrt{1-\overline{\alpha}_t}\boldsymbol{\epsilon}, t)\Vert^2\right]$$
\begin{equation}
\mathbf{x}_t(\mathbf{x}_0,\boldsymbol{\epsilon})=\sqrt{\overline{\alpha}_t}\mathbf{x}_0+\sqrt{1-\overline{\alpha}_t}\boldsymbol{\epsilon}
\label{equation:reparam1}
\end{equation}
\begin{equation}
    \boldsymbol{\mu}_\theta(\mathbf{x}_t,t)=\frac{1}{\sqrt{\alpha_t}}\left(\mathbf{x}_t-\frac{\beta_t}{\sqrt{1-\overline{\alpha}_t}}\boldsymbol{\epsilon}_\theta(\mathbf{x}_t,t)\right)
\label{eqaution:reparam2}
\end{equation}
The complete proof of these formulas can be found in \cite{ho2020denoising}.

\begin{table*}[th]
    \centering
    \caption{Overall performance comparison, including: (i) Mean Opinion Score (MOS) of speech expressiveness with 95\% confidence intervals; (ii) Jensen Shannon (JS) Divergence between the prosody distributions of the prediction and ground truth; (iii) Number of network parameters in the prosody predictors.}
    \label{tab:overall-perform}
    \begin{tabular}{cccccc}
        \toprule
        \multirow{2}{*}{\textbf{Systems}} & \multirow{2}{*}{\textbf{MOS ($\mathbf{\uparrow}$)}} & \multicolumn{3}{c}{\textbf{JS Divergence ($\mathbf{\downarrow}$)}} & \textbf{\#Params in}\\
         & & Pitch & Energy & Duration & \textbf{prosody predictor} \\
        \midrule
        FastSpeech2 with original prosody predictor  & $3.58 \pm 0.07$ & 0.199 & 0.056 & 0.119 & 1,185,027 \\
        FastSpeech2 with proposed prosody predictor  & $\mathbf{3.98 \pm 0.08}$ & $\mathbf{0.085}$ & $\mathbf{0.055}$ & $\mathbf{0.056}$ & $\mathbf{738,499}$ \\
        \midrule
        FastSpeech2 with ground truth prosody & $4.23 \pm 0.07$ & - & - & - & - \\
        Ground truth audio (Reconstructed) & $4.39 \pm 0.06$ & - & - & - & - \\
        \bottomrule
    \end{tabular}
\end{table*}
\begin{figure*}[th]

\begin{minipage}[t]{.33\linewidth}
  \centering
  \centerline{\includegraphics[width=5.9cm]{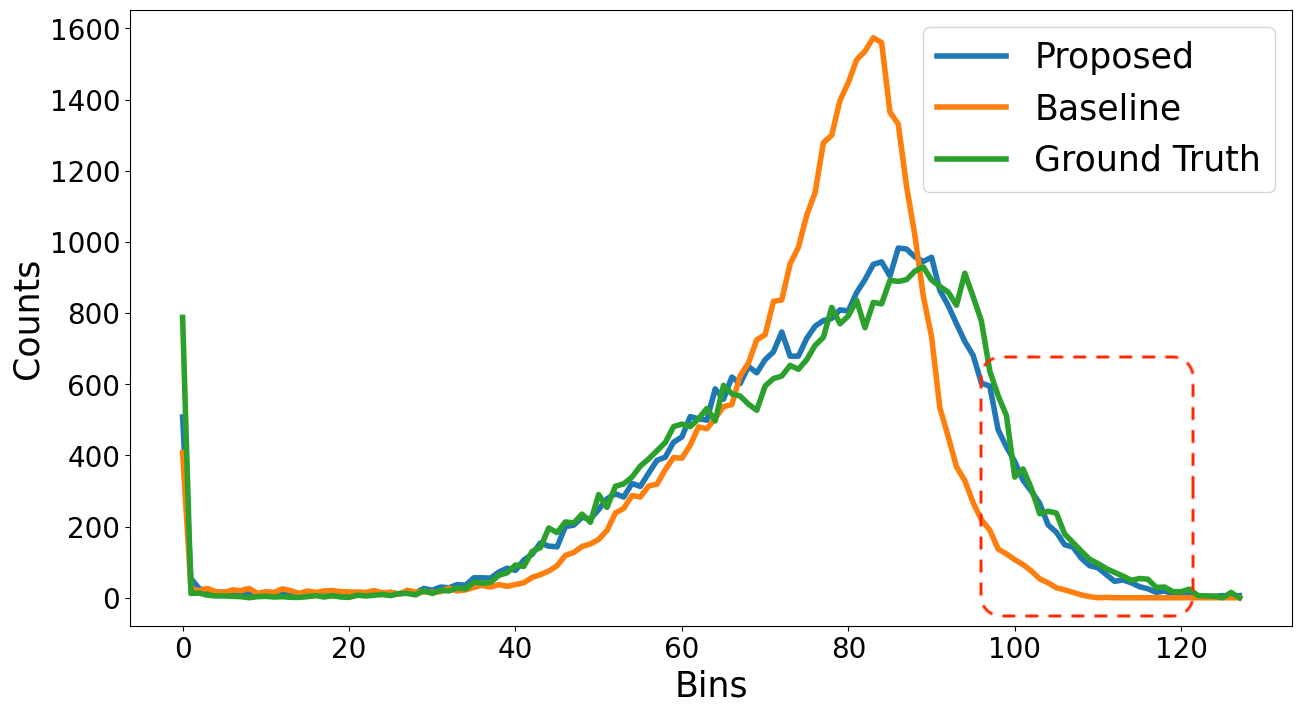}}
  \centerline{(a) Pitch}\medskip
\end{minipage}
\begin{minipage}[t]{.33\linewidth}
  \centering
  \centerline{\includegraphics[width=5.7cm]{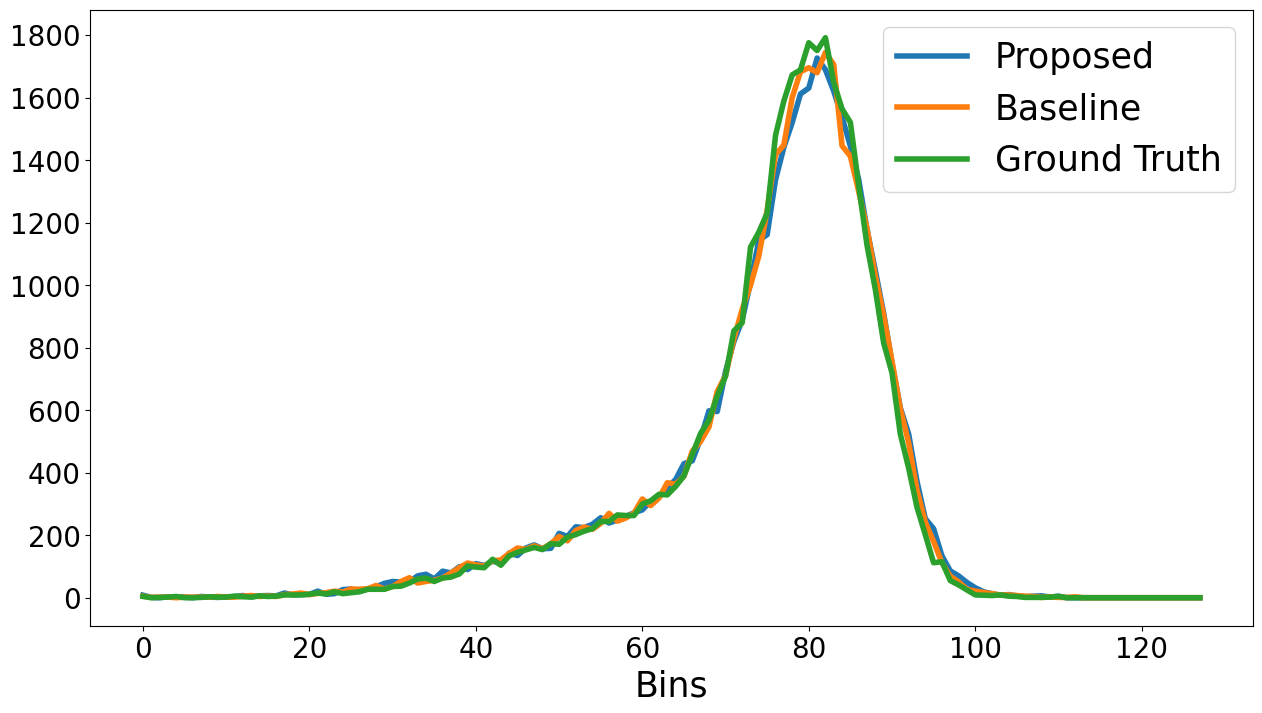}}
  \centerline{(b) Energy}\medskip
\end{minipage}
\hfill
\begin{minipage}[t]{.33\linewidth}
  \centering
  \centerline{\includegraphics[width=5.7cm]{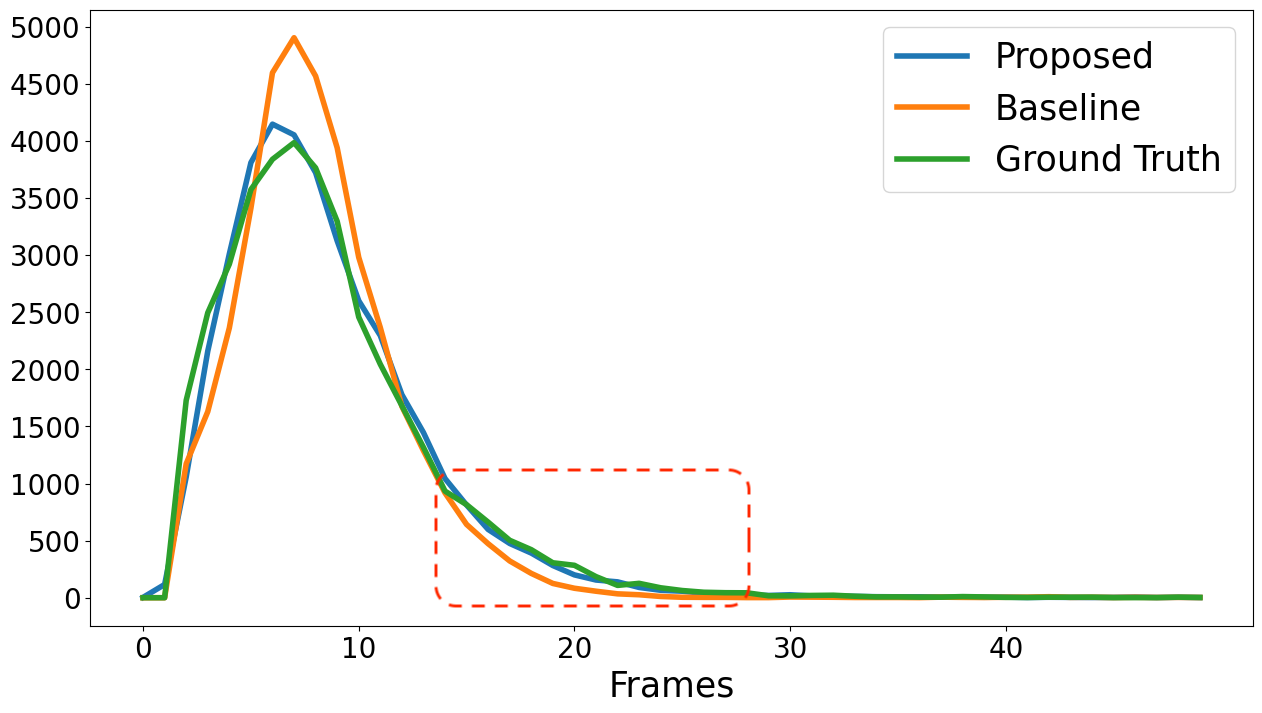}}
  \centerline{(c) Duration}\medskip
\end{minipage}
\caption{Distribution of the quantized prosody in predictions and ground truth for phoneme \textit{``ng''}. Red rectangles highlight that the baseline tends to avoid making aggressive predictions, resulting in over-smoothed distributions.
\textit{(Orange lines correspond to the baseline, blue lines correspond to the proposed method, green lines correspond to the ground truth distribution)}}
\label{fig:dist}
\end{figure*}

\subsubsection{DDPM-based prosody predictor}

The proposed prosody predictor is a DDPM modeled on 3-dimensional prosody features $\mathbf{x}_0$, which consists of phoneme-wise pitch, energy, and duration, respectively.
We employ non-causal WaveNet as the underlying denoiser network $\theta$ \cite{DBLP:conf/iclr/KongPHZC21,liu2022diffsinger}.

As shown in Figure~\ref{fig:diffvar_train}, during each training step of the prosody predictor,
we first take the text information generated by the FastSpeech2 encoder as prediction condition $c$.
Note that the parameters of the original FastSpeech2 modules are frozen the whole time to keep the TTS backbone intact.
Then we uniformly sample the Markov chain time step $t$ from $[1,T]$,
based on which the pre-extracted ground truth (GT) phoneme prosody features $\mathbf{x}_0$ are diffused into $\mathbf{x}_t$ via the reparameterization in Formula \ref{equation:reparam1}.
Given condition $c, t$ and input $\mathbf{x}_t$, the denoiser $\theta$ outputs $\boldsymbol{\epsilon}_\theta(\mathbf{x}_t,c, t)$, and updates its parameters by the gradient in Formula \ref{equation:grad}.
\begin{equation}
    \nabla_\theta\big\Vert\boldsymbol{\epsilon}-\boldsymbol{\epsilon}_\theta(\sqrt{\overline{\alpha}_t}\mathbf{x}_0+\sqrt{1-\overline{\alpha}_t}\boldsymbol{\epsilon},c, t)\big\Vert^2
    \label{equation:grad}
\end{equation}

The inference procedure of the prosody predictor is depicted in Figure~\ref{fig:diffvar_infer}.
Given text information $c$, we first sample the denoise start $\mathbf{x}_T\sim\mathcal{N}(\mathbf{0},\mathbf{I})$,
then iteratively compute $\mathbf{x}_{t-1}$ based on $\mathbf{x}_{t}$ according to Formula \ref{equation:eq2} for $t=T,\dots,1$, where $\mathbf{z}\sim\mathcal{N}(\mathbf{0},\mathbf{I})$ except for $\mathbf{z}=\mathbf{0}$ when $t=1$.
\begin{equation}
\mathbf{x}_{t-1}=\frac{1}{\sqrt{\alpha_t}}\left(\mathbf{x}_t-\frac{\beta_t}{\sqrt{1-\overline{\alpha}_t}}\boldsymbol{\epsilon}_\theta(\mathbf{x}_t,c,t)\right)+\sigma_t\mathbf{z}
  \label{equation:eq2}
\end{equation}
After looping for $T$ times, the final prosody prediction $\mathbf{x}_0$ is obtained. 
The $\mathbf{x}_0$ is subsequently sent to the modified variance adaptor in the TTS backbone and eventually guides the mel spectrogram decoder to synthesize speech stylized by the predicted prosody.
\begin{figure}[t]
  \centering
  \includegraphics[width=0.8\linewidth]{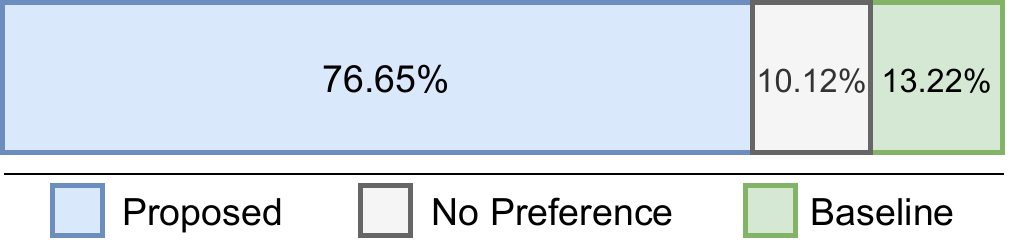}
  \caption{AB preference test results}
  \label{fig:abx}
  \vspace{-0.5cm}
\end{figure}
\begin{figure*}[th!]

\begin{minipage}[t]{.33\linewidth}
  \centering
  \centerline{\includegraphics[width=5.9cm]{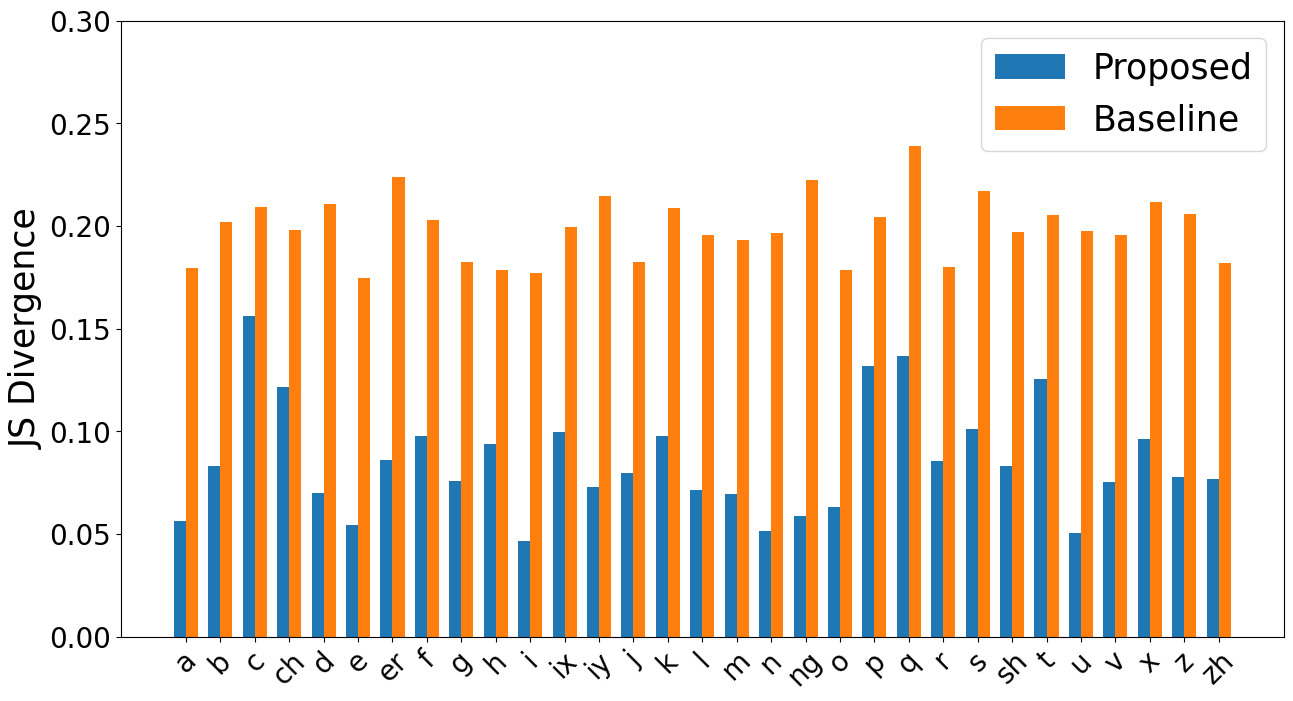}}
  \centerline{(a) Pitch}\medskip
\end{minipage}
\begin{minipage}[t]{.33\linewidth}
  \centering
  \centerline{\includegraphics[width=5.7cm]{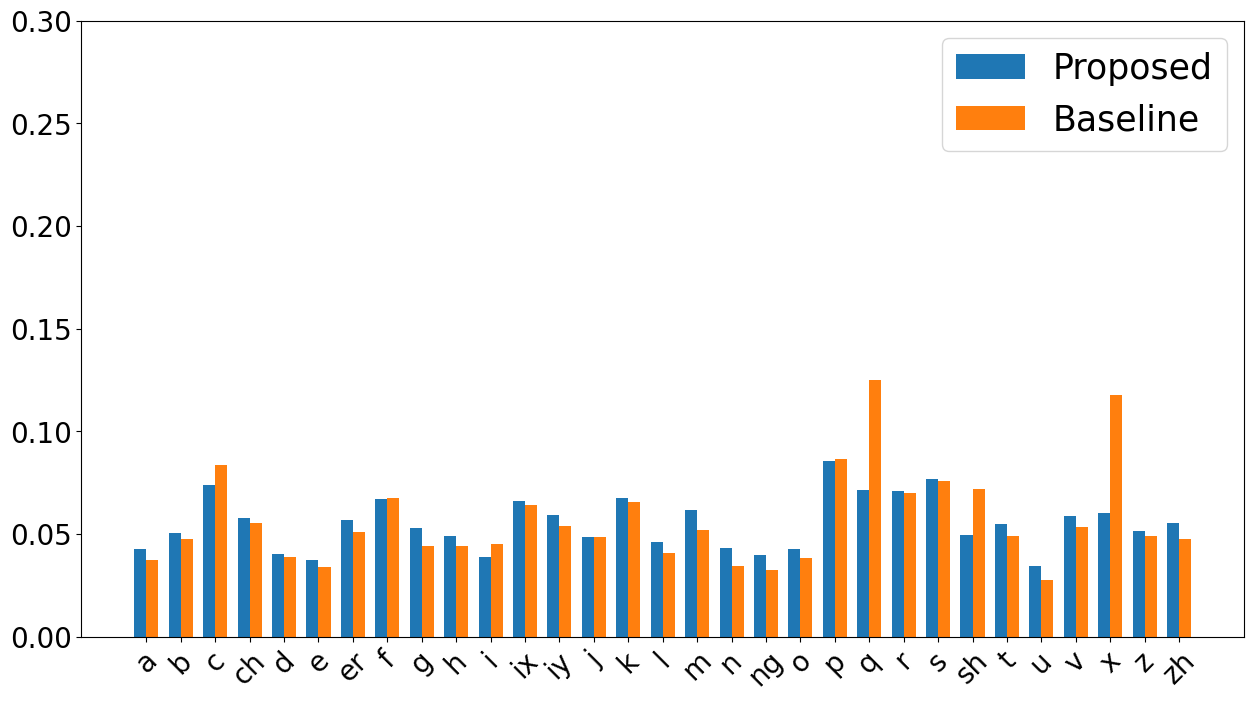}}
  \centerline{(b) Energy}\medskip
\end{minipage}
\hfill
\begin{minipage}[t]{.33\linewidth}
  \centering
  \centerline{\includegraphics[width=5.7cm]{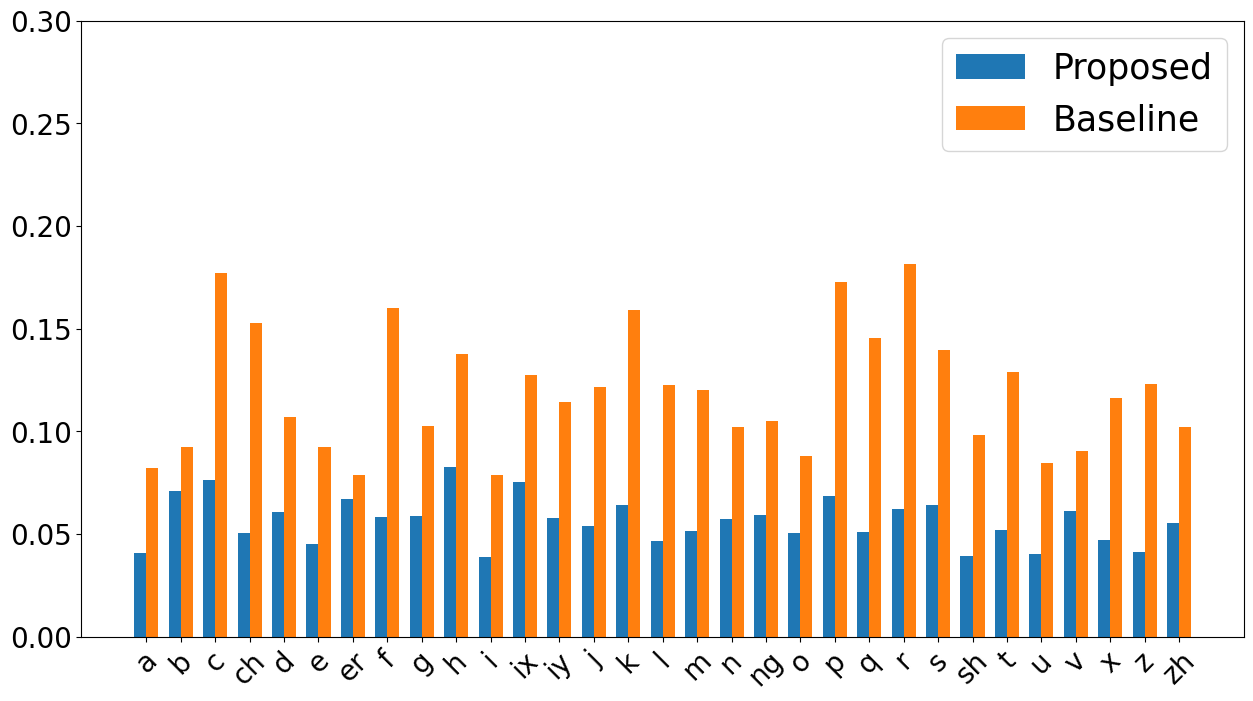}}
  \centerline{(c) Duration}\medskip
\end{minipage}
\caption{Detailed values of JS Divergence between the predicted and ground truth prosody over each phoneme. Higher bars stand for larger divergence, indicating worse diversity of the prediction. \textit{(Orange bars correspond to the baseline, blue bars correspond to the proposed method)}}
\label{fig:js_all}
\end{figure*}
\begin{figure}[th]

\begin{minipage}[t]{.5\linewidth}
  \centering
  \includegraphics[width=7.5cm]{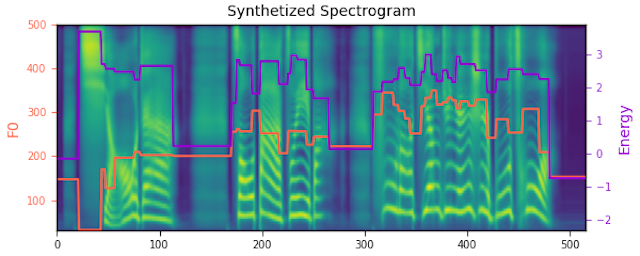}
\end{minipage}
%

\begin{minipage}[t]{.5\linewidth}
  \centering
  \includegraphics[width=7.1cm]{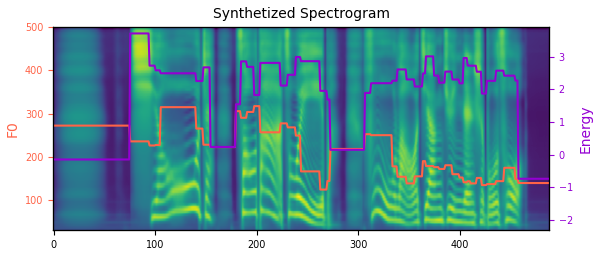}
\end{minipage}

\begin{minipage}[t]{.5\linewidth}
  \centering
  \includegraphics[width=6.3cm]{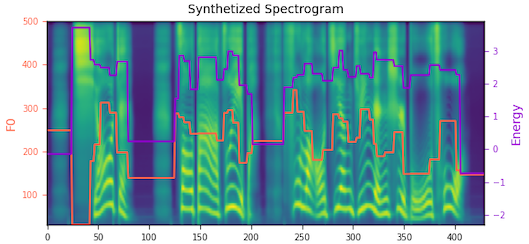}
\end{minipage}
\caption{Prosody prediction results generated by the proposed method in different runs on the same input text, stylizing the synthesized audio with diverse prosody.
}
\label{fig:spec}
\vspace{-0.5cm}
\end{figure}
\section{Experiments}
\label{sec:expr}

\subsection{Dataset and model details}
\label{sec:expr-data_model}
We apply the proposed method to a private Mandarin audiobook dataset with 28.23 hours of speech audio voiced by a professional male speaker,
since there is few public Mandarin corpus with diverse and expressive prosody variations, as well as satisfying sound quality.
We randomly sample 5\% of the dataset and reserve it for validation and testing, while the rest 95\% is used for training. 
All of the phoneme duration is extracted by a pre-trained speech recognition model automatically.
Based on which, the phoneme-wise pitch and energy features are obtained through averaging over frame-wise pitch and energy,
where the frame-wise pitch is extracted via Crepe \cite{kim2018crepe}, and the frame-wise energy is computed as the L2-norm of the amplitude of each short-time Fourier transform frame.
To boost the network's training, the prosody predictor's practical duration target is computed by taking the log scale of the raw phoneme duration.

The encoder and mel spectrogram decoder of the TTS backbone are made up of 4 and 6 Feed-Forward Transformer blocks \cite{ren2019fastspeech} with channel width 256, respectively.
The variance adaptor quantizes input pitch and energy values into 128 bins.
The proposed prosody predictor is built upon a DDPM with $T=500$ and linear noise schedule $\beta_{1:T}=\{10^{-4},\dots,0.06\}$,
and the underlying non-causal WaveNet $\theta$ consists of 10 stacked residual layers with channel width 64.
We also train the original prosody predictors in FastSpeech2 with mean squared error (MSE) as the deterministic \textbf{baseline} model for comparison,
which contains 60.5\% more network parameters, as shown in Table~\ref{tab:overall-perform}.
The baseline predictors are trained along with the FastSpeech2 backbone with batch size 16 on an Nvidia GeForce RTX 2080Ti for 900,000 steps.
The proposed model is trained under the same setting for the same number of steps, which took approximately 31 hours.
The training and evaluation of both methods are performed in a single run, without grid searching for hyperparameters and random seeds.
A pre-trained HiFi-GAN \cite{DBLP:conf/nips/KongKB20} is utilized as the backend for waveform reconstruction.

\subsection{Expressiveness}
\label{sec:expr-expressiveness}
To evaluate the expressiveness of the speech synthesized by the proposed method, we conduct a Mean Opinion Score (MOS) test and an AB preference test on 22 utterances from the test set, with 22 native speakers serving as subjects.
In the MOS test, each subject is asked to rate the given audios on a scale of 1 to 5 regarding their expressiveness.
As shown in Table~\ref{tab:overall-perform},
the proposed method reaches higher scores than the baseline.
In the AB test, each subject is asked to choose the more expressive audio of 2 anonymous ones synthesized by the proposed method and baseline.
As shown in Figure~\ref{fig:abx}, the proposed method is favored with an overwhelming 76.65\% of the preference, while the baseline only receives 13.22\%.

The prevailing subjective test results on expressiveness are consistent with the statistics of the prosody prediction distribution.
As plotted in Figure~\ref{fig:dist}, using the phoneme \textit{``ng''} as an example,
the predictions from the proposed method accumulate to a distribution closer to ground truth distribution.
In contrast, the baseline tends to make over-smoothed predictions, resulting in squeezed distributions.
From our empirical study,
the ground truth distributions of the pitch and duration are more complex than that of the energy,
which appears to have low variance in a narrow bell curve.
This explains the defective performance of the baseline,
as it is limited in unimodal predictions.

\subsection{Diversity}
\label{sec:expr-diversity}
To reveal the diversity of the predicted prosody, we compute the Jensen Shannon (JS) divergence between the distributions of the quantized prosody in predictions and ground truth
to measure how well the predictions fit into the ground truth distribution \cite{huang2022prodiff}.
Results in Table~\ref{tab:overall-perform} show that the proposed prosody predictor achieves lower JS divergence on all of the 3 prosody features, indicating a better diversity of prediction results.
We also plot out the JS divergence values over each phoneme in Figure~\ref{fig:js_all} to observe the characteristics of prediction diversity, which turns out to match the observation in Figure~\ref{fig:dist}.
For pitch and duration prediction, the proposed method produces better results on all the phonemes.
For energy prediction, there is no significant difference on most phonemes except a few voiceless sounds like \textit{``q''}, \textit{``x''} and  \textit{``sh''}.

We can also directly demonstrate the diversity of the proposed method by sampling multiple times on the same text.
As shown in Figure~\ref{fig:spec}, the predicted prosody varies in different samples,
resulting in diverse synthesized spectrograms.
Meanwhile, the predicted prosody is still stably in coherence with the text, as reflected in Table~\ref{tab:overall-perform} where no significant expansion of confidence interval in the MOS result
is observed.
\section{Conclusion and Discussion}
\label{sec:method}
In this work, we propose a DDPM-based speech prosody predictor and attach it to a FastSpeech2-based TTS backbone to form an expressive TTS system.
Compared with traditional deterministic predictors, the proposed DDPM-based method enhances the expressiveness of synthesized speech and brings noticeable improvements to the diversity of the predicted prosody,
which is still in coherence with the given text.
Please refer to our demo website\footnote{https://diffvar.github.io/DDPM-prosody-predictor} for listening to samples.

Nevertheless, the iterative sampling process of the proposed method brings increased latency to the TTS system.
We conduct a computation cost test on an Nvidia GeForce RTX
2080Ti, and find that the average real-time factor (RTF) of the proposed
DDPM-based prosody predictor with full 500 denoising steps is $0.47$, while the RTF of the original FastSpeech2 variance predictor is $0.31\times 10^{-3}$.
Although the DDPM-based approach achieves real-time inference, it is slowed down by the iterative sampling process.
In the future, we will consider introducing fast sampling techniques that are able to produce results within a couple of denoising steps,
as well as
modifying the underlying DDPM with novel improvements that emerge in the latest diffusion models \cite{salimans2022progressive,lam2022bddm}.

\section{Acknowledgements}
This work is supported by National Natural Science Foundation of China (62076144), Shenzhen Science and Technology Program (WDZC20220816140515001), Tencent AI Lab Rhino-Bird Focused Research Program (RBFR2022005) and Tsinghua University - Tencent Joint Laboratory.



\bibliographystyle{IEEEtran}
\bibliography{mybib}

\end{document}